# AI in the Loop - Functionalizing Fold Performance Disagreement to Monitor Automated Medical Image Segmentation Pipelines

Harrison C. Gottlich, B.A.[1*], Panagiotis Korfiatis, Ph.D[2*], Adriana V. Gregory, M.S.[2], Timothy L. Kline, Ph.D.[2,3,#]

[1]Mayo Clinic Alix School of Medicine, Mayo Clinic, Rochester, MN, USA

[2]Department of Radiology, Mayo Clinic, Rochester, MN, USA

[3]Division of Nephrology and Hypertension, Mayo Clinic, Rochester, MN, USA

[*] Shared co-first authorship

**[#] Corresponding author:**
Timothy Kline, Ph.D.
Telephone: 507-284-2804
Email: Kline.Timothy@mayo.edu
Address: 200 First St SW, Rochester, MN 55905, USA

**Abstract**

Methods for automatically flag poor performing-predictions are essential for safely implementing machine learning workflows into clinical practice and for identifying difficult cases during model training. We present a readily adoptable method using sub-models trained on different dataset folds, where their disagreement serves as a surrogate for model confidence. Thresholds informed by human interobserver values were used to determine whether a final ensemble model prediction would require manual review. In two different datasets (abdominal CT and MR predicting kidney tumors), our framework effectively identified low performing automated segmentations. Flagging images with a minimum Interfold test Dice score below human interobserver variability maximized the number of flagged images while ensuring maximum ensemble test Dice. When our internally trained model was applied to an external publicly available dataset (KiTS21), flagged images included smaller tumors than those observed in our internally trained dataset, demonstrating the methods robustness to flagging poor performing out-of-distribution input data. Comparing interfold sub-model disagreement against human interobserver values is an efficient way to approximate a model's epistemic uncertainty – its lack of knowledge due to insufficient relevant training data – a key functionality for adopting these applications in clinical practice.

**Introduction**

Automated medical image segmentation techniques have numerous benefits for healthcare delivery. Deep learning-based image segmentations have demonstrated applications across many body areas and imaging modalities (Hesamian et al., 2019; Siddique et al., 2021; Yagi et al., 2019; Zhao et al., 2018), being used directly to measure organ volume or for 3D modeling and printing to help patients understand the anatomical basis of their disease as well as to educate surgical trainees through high-fidelity simulations (Rickman et al., 2021). Automated segmentations can also be one part of a workflow where predictions feed into additional models to classify pathology and to inform medical decision-making. However, a challenge of implementing such a workflow is ensuring robust quality control of automated segmentations without the potentially infeasible burden of continuous human monitoring (Alakwaa et al., 2017).

Given the potential value of medical image segmentation, the topic of model development has been intensely studied on internal datasets as well as in open-sourced challenges (Bilic et al., 2019; Heller et al., 2021; Heo, 2022). Recently, the "no new U-Net model" (nnU-Net) framework has consistently produced winning submissions in a number of these open challenges (Isensee et al., 2021). nnU-Net automates many neural network design choices, allowing researchers to focus on other barriers for clinical implementation. One significant barrier is the possibility of models encountering data not represented well in the training set, otherwise referred to as "out-of-distribution" data, due to original training data bias, task data drift, or unique scenarios. Models in the U-Net family will always generate predictions on properly formatted input data without a measure of certainty in its output, potentially leading to catastrophic consequences in research or clinical decision-making unless oversight of automated processes exists. For instance, a model trained on cross-sectional (CT) images would still yield a prediction if magnetic resonance (MR) data was accidentally used, despite not having seen MR image data before. The potential mismatch between training data and task data is a major issue for clinical implementation of automated deep learning models unable to flag problematic predictions (Shaw et al., 2019; Vayena et al., 2018). This lack-of-knowledge of the models own limitations due to limited training data is also referred to as epistemic uncertainty (Ghoshal et al., 2021; Lakshminarayanan et al., 2017; Swiler et al., 2009).

Several approaches have been proposed to address the issue of out-of-distribution task data in clinical implementations of AI workflows (Abdar et al., 2021; Ghoshal et al., 2021; Lakshminarayanan et al., 2017; Zheng et al., 2022). These methods are a form of "AI in the Loop" where separate automated model processes are inserted into workflows to automatically check predictions and flag instances when a human intervention may be needed. While not replacing the need for a representative dataset and well validated models, these methods can provide an important safeguard to protect research interests and patient care when neither a reference standard nor human oversight is feasible for each prediction.

In this paper, we propose an easily implementable framework that equips conventionally trained 5-fold-cross validation models with the ability to monitor real-time predictions when reference standards are unavailable. This AI-in-the-Loop method is novel, being easily understandable, quickly computable, and powerfully enabling a clinically implemented image segmentation workflow to discriminate when a prediction segmentation requires human review.

## Materials and Method

### Dataset

This multi-dataset retrospective study was approved by our institutional review board, was HIPAA compliant, and performed in accordance with the ethical standards contained in the 1964 Declaration of Helsinki. We used two internal datasets: 1) a MR abdomen dataset with labeled Kidney and Tumor and 2) a CT abdomen dataset with labeled Kidney and Tumor. Additionally, the open-sourced KiTS21 dataset as described in the publication by Heo et al (Heller et al., 2020; Heo, 2022). was used to demonstrate out-of-distribution task data. Our internal datasets are described in detail with demographic data in **Table 1**.

**Table 1. Internal Dataset Demographics**

|  | **CT dataset** | **MR dataset** |
|---|---|---|
| No. of subjects | 350 | 350 |
| Males | 229 | 217 |
| Females | 121 | 133 |
| *Age | 63 ± 13 [19 – 88] | 59 ± 14 [20 – 88] |
| *Height (m$^2$) | 1.72 ± 0.1 [1.43 – 2.04] | 1.73 ± 0.1 [1.49 – 2.04] |
| *Weight (Kg) | 92.79 ± 25.02 [45 – 200] | 90.17 ± 22.33 [46 – 190] |
| *BMI (Kg/m$^2$) | 31.00 ± 7.34 [16 – 62] | 30.13 ± 6.52 [17 – 57] |

* Mean ± standard deviation

**MR Kidney/Tumor Dataset**

As part of a previously published studied, 350 T2-weighted with fat-saturation coronal MR abdominal/pelvis images were randomly sampled from a dataset of 501 patients where 313 underwent partial nephrectomy and 188 underwent radical nephrectomy between 1997 and 2014 (Gottlich et al., 2023). Segmentation of these images was first performed by a a previously trained U-Net based algorithm predicting labels for right and left kidney (Kline et al., 2017; van Gastel et al., 2019). Then, two urologic oncology fellows manually refined these automatic segmentations and segmented renal tumors as well.

**CT Kidney/Tumor Dataset**

350 images were randomly sampled from a collection of 1,233 non-contrast and different contrast phase abdomen/pelvis CT images as part of the Mayo Clinic Nephrectomy Registry (Denic et al., 2020). Images were from 2000-2017 in patients without metastatic lesions or positive lymph nodes at time of radical nephrectomy. Two urologic oncology fellows segmented kidney and tumor masks using the segmentation software ITK-snap **RRID:SCR_002010** (version 2.2; University of Pennsylvania, Philadelphia, PA) (Yushkevich et al., 2006). Processing of these images included cropping around both kidneys and 3 slices above the slice of the upper pole of the kidney and 3 slices below the lower pole of the kidney. Also, the scans were resampled to a 256-pixel coronal plane width and 128-pixel medial plane depth with zero padding employed if images were smaller than this standard size.

**Algorithm**

**nnU-Net Specifications**

nnU-Net preprocessing consisted of designating for each dataset either "T2" or "CT" default processing. Following the public GitHub **RRID:SCR_00263** (Beck et al., 2017), standard five-fold cross validation using 3d_fullres was used where final predictions are derived by averaging the 5 sub-model output voxel-wise softmax probabilities into one ensemble prediction (Isensee et al., 2021). In addition, each sub-model prediction was also evaluated to assess fold disagreement.

**Self-informed models**

Our main goal is to utilize the information encoded in models generated during the 5-fold cross validation process and investigate if information extracted during the inference stage can inform the end user of the final ensemble model segmentation quality.

During training, a 5-fold cross validation approach was utilized leading to the generation of 5 sub-models. In this paper we define a sub-model as a fully trained model that has a unique training and validation test set split. Given a test image the predictions are ensembled utilizing softmax probabilities, where the probabilities across folds are averaged and to determine the final prediction. For each test image, we calculated Dice scores between each sub-model prediction i.e. Dice between sub-model 1 and sub-model 2, sub-model 1 and sub-model 3, and so on. Dice score is a commonly used metric to compare 3D image segmentations, where a score of 1 indicates complete overlap between the two segmentations and 0 is of two segmentations with no overlap (Taha and Hanbury, 2015). This process produced 10 Dice metrics referred to as "Interfold Dices" that were summarized by different first order summary statistics to compare against published human-human interobserver thresholds. As part of our investigations into metrics for flagging cases where the ensemble model's prediction may be suboptimal, we evaluated the following first order statistics: mean, median, min, and max of the Dice index.

To validate our method, we compared the summarized Interfold Dice with the final test ensemble Dice score to investigate whether an association existed. We first created scatter plots where the y-axis was a given summary metric of the Interfold Dice, and the x-axis was the eventual ensemble model Dice score. Intuitively, we also used confusion matrices to display the results, where true positives were flagged images based on summary of Interfold Dice with poor ensemble model performance, true negatives were non-flagged segmentations with good ensemble model performance, false positives were flagged images with good ensemble model performance, and false negatives were non-flagged images with poor ensemble model performance. Of these categories, false negatives were considered the worst failure since they represented unflagged poor segmentations that might not be reviewed before being used in a clinical workflow. False positives were undesirable but not overtly worrisome in small quantities since they represent flagged images that had good performance and could be quickly reviewed. We also calculated how the overall test Dice set score would change if the flagged segmentations were removed.

**Simulating Out-of-Distribution Task Data**

To test the generalizability of our framework in identifying "out-of-distribution " data, we used our internally trained model to predict segmentations on the open-sourced KiTS21 dataset (Heller et al., 2020) knowing that key differences existed between the datasets. The external CT dataset consists of images acquired from contrast-enhanced CT scans during the corticomedullary contrast phase and consists of generally smaller tumors including those from partial nephrectomies. In contrast, the internal CT dataset contains a mix of different contrast phases and larger tumors, being sourced solely from a radical nephrectomy database. The difference in voxel dimensions and distribution of tumor size between the KiTS21 dataset and our internal dataset can be found in **Table 2.**

**Table 2. Dataset Voxel and Volume Characteristics**

|  | **MR Kidney Tumor** | **CT Kidney Tumor** | **KiTS21** |
|---|---|---|---|
| **In-plane voxel width x height (mm)** |  |  |  |
| Mean | $1.34 \times 1.34$ | $1.03 \times 1.03$ | $0.79 \times 0.79$ |
| Median | $1.56 \times 1.56$ | $1.01 \times 1.01$ | $0.78 \times 0.78$ |
| Range | $0.59 - 1.95 \times 0.59 - 1.95$ | $0.49 - 1.85 \times 0.49 - 1.85$ | $0.44 - 1.04 \times 0.44 - 1.04$ |
| **Slice thickness (mm)** |  |  |  |
| Mean ± Std | $6.26 \pm 1.65$ | $4.03 \pm 1.38$ | $3.18 \times 1.75$ |
| Median | 6.00 | 5.00 | 3.00 |
| Range | 2.00 – 15.00 | 0.65 – 8.00 | 0.50 – 5.00 |
| **Number of Slices** | | | |
| Mean ± Std | 32.27 ± 11.64 | 44.50 ± 23.91 | 314.67 ± 37.83 |
| Median | 30.00 | 37.00 | 320.00 |
| Range | 6.00 – 116.00 | 20.00 – 211.00 | 257.00 - 478.00 |
| **Volume of Labels (ml)** | | | |
| Tumor minimum volume | 0.01 | 0.513 | 1.86 |
| Tumor 25th percentile volume | 29.61 | 55.95 | 18.71 |
| Tumor Mean ± Std | 541.33 ± 1244.35 | 428.61 ± 654.27 | 253.59 ± 476.38 |
| Tumor Median volume | 97.33 | 212.71 | 66.77 |
| Tumor 75th percentile volume | 454.70 | 497.67 | 219.46 |
| Tumor maximum volume | 11,946.84 | 7,742.26 | 2894.02 |

## Results

### MR Kidney/Tumor Results

Interobserver Dice for MR tumor labels has been previously found to be 0.825 on average (Rasmussen et al., 2022). In our study, the ensemble model performance without flagging for MR Kidney tumor was 0.76 ± 0.27. The full results of the impact of flagging with different summary metrics can be found below in **Table 3** and **Figures 1a-1b.** All unflagged cohorts mean ensemble Dice values were above the human interobserver value with small standard deviations.

**Table 3. Different Summary of Interfold Dice Flag Cohort – MR Kidney Tumor**

| Summary Metrics | Number of Flagged Images | Flagged Images Mean ± Standard Deviation | Unflagged Images Mean ± Standard Deviation |
|---|---|---|---|
| Mean | 17 | 0.53 ± 0.37 | 0.87 ± 0.07 |
| Median | 8 | 0.28 ± 0.36 | 0.85 ± 0.09 |
| Max | 4 | 0 ± 0 | 0.82 ± 0.15 |
| Min | 20 | 0.57 ± 0.36 | 0.87 ± 0.08 |

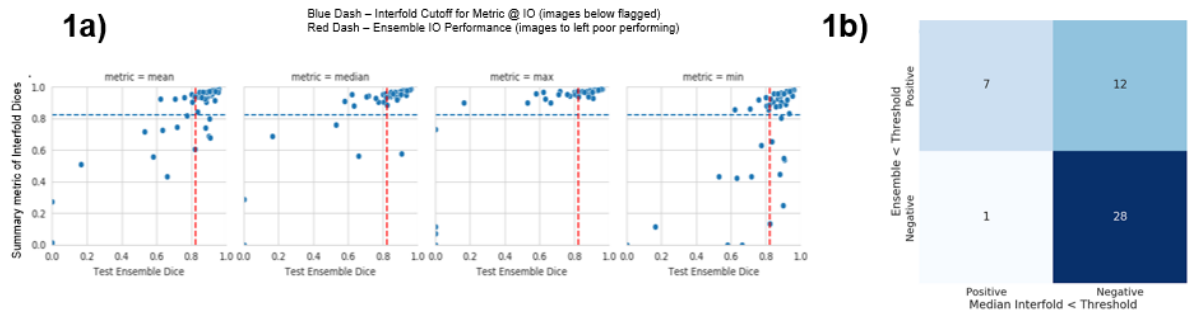

**Figure 1a and 1b. MR Kidney Tumor Characteristics of Flagged and Non-Flagged Images.** True positive is defined as when flagged images (summary Interfold Dice < threshold) performed poorly (test ensemble < threshold). True Negative is defined as when non-flagged images (summary Interfold Dice > threshold) performed well (test ensemble > threshold). False positives defined as when flagged images (summary Interfold Dice < threshold) performed well (test ensemble > threshold). False negatives defined as when non-flagged images (summary Interfold Dice > threshold) performed poorly (test ensemble < threshold).

**CT Kidney/Tumor Results**

Interobserver Dice values for CT Kidney Tumor has been previously found to be 0.90 on average (Rasmussen et al., 2022). The mean ensemble Dice model performance for CT Kidney was mean ± standard deviation was 0.93 ± 0.02. The full results of the impact of flagging with different summary metrics can be found below in **Table 4** and **Figures 2a-2d.** Almost all unflagged cohorts mean ensemble Dice values were above the human interobserver value with small standard deviations.

**Table 4. Different Summary of Interfold Dice Flag Cohort – MR Kidney Tumor**

| Summary Metrics | Number of Flagged Images | Flagged Images Mean ± Standard Deviation | Unflagged Images Mean ± Standard Deviation |
|---|---|---|---|
| Mean | 9 | 0.57 ± 0.35 | 0.91 ± 0.08 |
| Median | 8 | 0.52 ± 0.34 | 0.91 ± 0.08 |
| Max | 4 | 0.32 ± 0.37 | 0.89 ± 0.10 |
| Min | 12 | 0.61 ± 0.31 | 0.92 ± 0.05 |

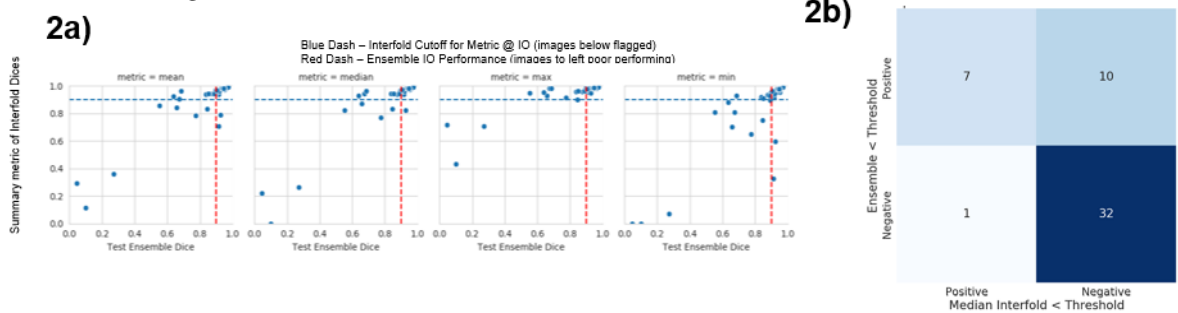

**Figure 2a and 2b.** True positive is defined as when flagged images (summary Interfold Dice < threshold) performed poorly (test ensemble < threshold). True Negative is defined as when non-flagged images (summary Interfold Dice > threshold) performed well (test ensemble > threshold). False positives defined as when flagged images (summary Interfold Dice < threshold) performed well (test ensemble > threshold). False negatives defined as when non-flagged images (summary Interfold Dice > threshold) performed poorly (test ensemble < threshold).

**Predictions on KiTS21 Results**

We applied the same comparison human interobserver CT Kidney and Tumor values to evaluate the internally trained model on the KiTS21 dataset. The mean test Dice for tumor was 0.67 ± 0.36 with significant improvement after removing the flagged cohort. The full results of the impact of flagging with different summary metrics can be found below in **Table 5** and **Figures 3a-3b.** All unflagged cohorts mean ensemble Dice were near the human interobserver value with small standard deviations.

**Table 5. Different Summary of Interfold Dice Flag Cohort – CT Kidney Tumor model on KiTS21 Data**

| Summary Metrics | Number of Flagged Images | Flagged Images Mean ± Standard Deviation | Unflagged Images Mean ± Standard Deviation |
|---|---|---|---|
| Mean | 143 | 0.37 ± 0.38 | 0.89 ± 0.11 |
| Median | 135 | 0.34 ± 0.37 | 0.89 ± 0.10 |
| Max | 87 | 0.13 ± 0.24 | 0.85 ± 0.17 |
| Min | 175 | 0.45 ± 0.39 | 0.91 ± 0.07 |

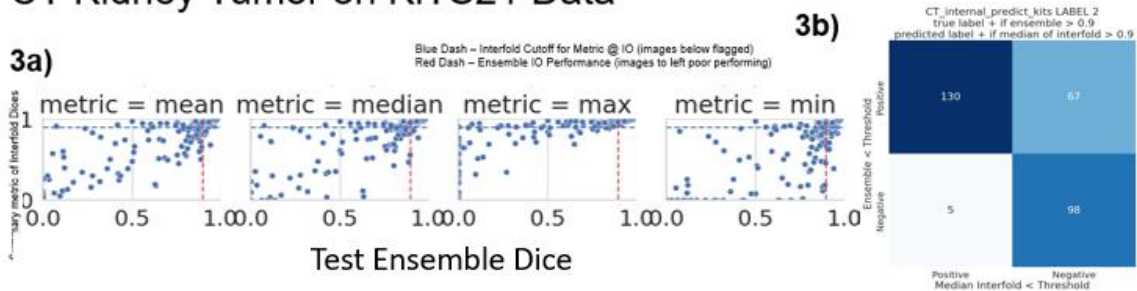

**Figure 3a and 3b.** True positive is defined as when flagged images (summary Interfold Dice < threshold) performed poorly (test ensemble < threshold). True Negative is defined as when non-flagged images (summary Interfold Dice > threshold) performed well (test ensemble > threshold). False positives defined as when flagged images (summary Interfold Dice < threshold) performed well (test ensemble > threshold). False negatives defined as when non-flagged images (summary Interfold Dice > threshold) performed poorly (test ensemble < threshold)

As seen in **Figure 4a-4d,** the flagged images tended to be of tumors smaller than observed in the models training set:

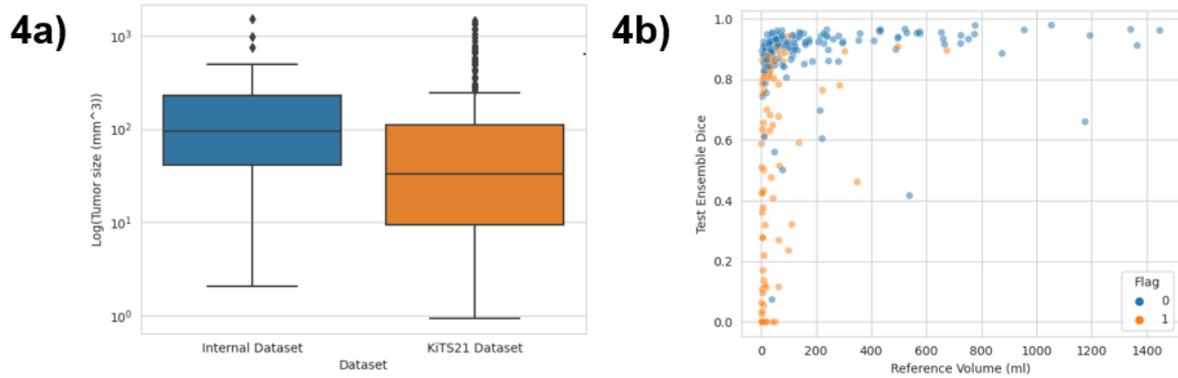

**Figure 4a and 4b.** Distribution of Tumor Sizes in internally trained dataset vs. KiTS21 with showing which KiTS21 Data are flagged. 4a) Boxplot graph demonstrating different tumor size distributions in CT datasets while 4b) demonstrates how flagged images tended to be smaller tumor volumes.

**Qualitative Assessment of Flagged Images**

In addition to how out-of-distribution tumor size affected whether the model would have higher epistemic uncertainty and the impact on final ensemble test performance, we also qualitatively assessed flagged outliers. An important finding for the CT Kidney and Tumor internal data test set is that outliers tended to represent more difficult segmentation cases as opposed to corrupted images as seen below in **Figure 6**.

**Figure 6.** Qualitative assessment of outliers in internal CT Tumor Test Set shown in lowest left quadrant of **Figure 3C**.

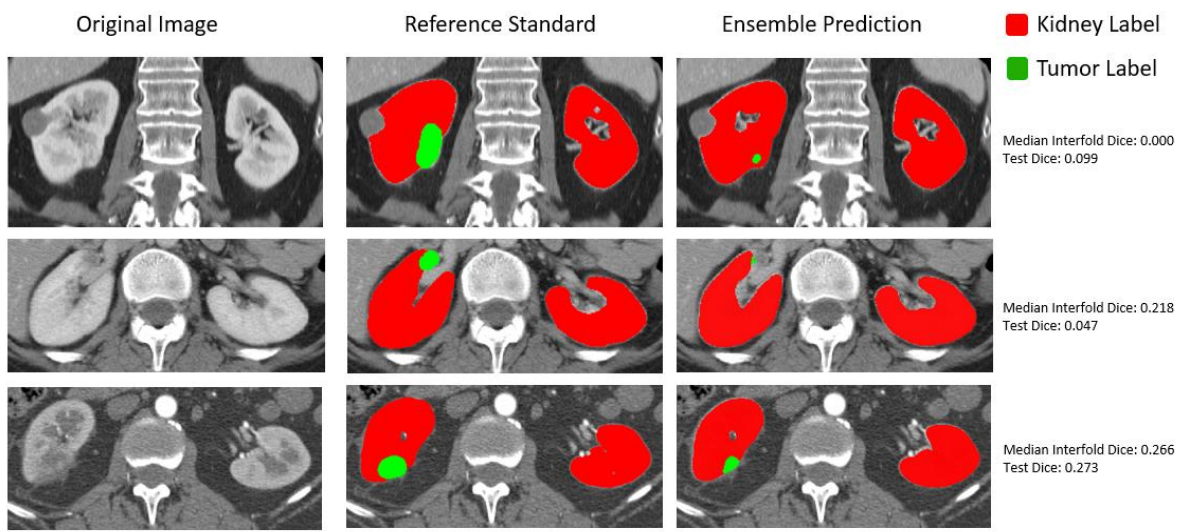

**Discussion**

The primary goal of this paper was to leverage a state-of-the-art convolutional neural network framework to create a self-informed model capable of informing users about the segmentation quality without requiring a reference standard. For this purpose, we utilized information from the five-fold cross validation steps that approximate epistemic uncertainty by identifying poorer inference performance related to a given task example being out-of-distribution of the model's training data.

An intuitive understanding of why this method works is based on how cross-validation uses different distributions in training and validation folds to minimize overfitting on a single distribution. Despite seeing different distributions of data, we still expect predictions from different folds to resemble one another if the test data distribution is well represented throughout the training and validation sets. However, in out-of-distribution or near out-of-distribution cases, we anticipate greater variance between folds depending on the split of limited relevant data between the training and validation cases. This variance reveals that the folds lack of adequate examples resulting in a less confident prediction and, as demonstrated in the above trials, lower final ensemble model performance. This application complements the importance of collating representative datasets and well validated models as an additional tool for clinically implemented image segmentation workflows to efficiently identify when a prediction may be poor due to a lack of relevant training data.

To identify poor performing predictions, we compared sub-model predictions to each other and summarized them with different metrics to obtain a single Interfold Dice score. This score was compared against published human interobserver thresholds to determine which images should be flagged in our hypothetical workflow. For tumor segmentation tasks, flagged images tended to be the poorest performing, while unflagged predictions had significantly higher mean Dice with less variability. Moreover, we demonstrated by applying our internal model to the KiTS21 dataset that, despite overall poor model performance, the unflagged cohort still performed comparable to human interobserver values, while the images in the flagged cohort were generally of a smaller tumor size distribution than observed in the training dataset.

This application is a contribution to addressing identified issues of automated medical image segmentation models, particularly by identifying when an algorithm is being applied to out-of-distribution data. Past quantitative work to detect out-of-distribution task data include creating separate classification models to identify out-of-distribution data and quantifying uncertainty using Markov chain Monte Carlo methods (Ghoshal et al., 2021; Zheng et al., 2022). Lakshminarayanan et al. published a method most similar to the one presented here in comparing ensemble models combined with adversarial training to identify out-of-distribution examples (Lakshminarayanan et al., 2017). Our study builds on this work by demonstrating a way to implement out-of-distribution detection in a medical image workflow by using human interobserver values as cut-offs for flagging predictions. This real-time monitoring not only offers workflow implementers the ability to correct flagged examples but also alerts them to investigate and identify causes out-of-distribution data. In some cases, such data may be corrupted input data or might represent a scenario requiring model updates (e.g., data drift scenarios requiring continuous learning or other model update paradigms).

A key limitation of this method is its inability to correct for poor in-distribution training data. For example, the model may make a poor prediction with high certainty based on the training data that it has been provided. This problem is especially important to address in terms of entrenched biases that might be present in datasets (Braveman, 2006; Vayena et al., 2018).

Regarding future directions, we plan to investigate methods for identifying less obvious causes of higher epistemic uncertainty. Furthermore, we believe that a prospective validation study demonstrating the method in real-time is crucial for assessing its suitability for clinical implementation. Lastly, we have made our analysis code open-sourced and easily accessible in an interactive manner for other investigators to determine its utility in different applications at the following link: https://colab.research.google.com/drive/1E4JBpl5X_9BXz_2AHUSM11z59CXD5DIc?usp=sharing.

## **Conclusions**

Evaluating interfold sub-model predictions offers a power and efficient approach to identifying epistemic uncertainty in segmentation models, a critical aspect for successful integration of these applications in clinical practice.

**Acknowledgements:**

The authors thank Naoki Takahashi, M.D., Andrew D. Rule, M.D., Aleksandar Denic, M.D. Ph.D., Vidit Sharma, M.D., and Abhinav Khanna, M.D. for their help with data curation and annotations.

We also acknowledge arXiv preprint submission system to allow us to share our work for public comment as we submit it for peer-review and consideration by journals.

**Funding:** This work was supported by Mayo Clinic and nference AI Challenge Award from Mayo Clinic Ventures, nference and the Mayo Clinic Office of Translation to Practice. Research reported in this publication was supported by the National Institute of Diabetes and Digestive and Kidney Diseases of the National Institutes of Health under Award Numbers K01DK110136, and R03DK125632. The content is solely the responsibility of the authors and does not necessarily represent the official views of Mayo Clinic or nference or the National Institutes of Health.